\documentclass[journal]{IEEEtran}

\usepackage{xcolor,soul,framed} 
\colorlet{shadecolor}{yellow}
\usepackage[pdftex]{graphicx}
\graphicspath{{../pdf/}{../jpeg/}}
\DeclareGraphicsExtensions{.pdf,.jpeg,.png}

\usepackage[cmex10]{amsmath}
\usepackage{array}
\usepackage{mdwmath}
\usepackage{mdwtab}
\usepackage{eqparbox}
\usepackage[hyphens]{url}

\begin{document}
\bstctlcite{IEEEexample:BSTcontrol}
    \title{Supply Chain Characteristics as Predictors of Cyber Risk: A Machine-Learning Assessment}
  \author{Kevin~Hu,
      Retsef~Levi,
      Raphael~Yahalom,
      and~El Ghali~Zerhouni

}

\markboth{
}{Roberg \MakeLowercase{\textit{et al.}}: High-Efficiency Diode and Transistor Rectifiers}

\maketitle

\begin{abstract}
This paper provides the first large-scale data-driven analysis to evaluate the predictive power of digital supply chain attributes for assessing risk of cyberattack data breaches. Motivated by rapid increase in the complexity of digital supply chains and related third-party enabled cyberattacks, the paper provides the first quantitative empirical evidence that digital supply-chain attributes are significant predictors of enterprise cyber risk.  The analysis leverages externally observable cybersecurity ratings that aim to capture the quality of the enterprise internal cybersecurity management, but augments these with original supply chain  features that are inspired by observed third-party cyberattack scenarios, as well as concepts from network science research. The main quantitative result of the paper is to show that these supply chain network features add significant detection power to predicting enterprise cyber risk, relative to merely using enterprise-only attributes. In particular, compared to a base model that relies only on internal enterprise features, the supply chain network features improve the out-of-sample AUC by 2.3\%.  Given that a cyber data breach on a specific enterprise is a low probability high impact risk event, these improvements in the prediction power have significant value. Additionally, the model highlights several cybersecurity risk drivers related to third-party cyberattack and breach mechanisms and provides important insights as to what key metrics should be monitored and what interventions might be effective to mitigate these risks. 
\end{abstract}


%
\IEEEpeerreviewmaketitle


\section{Introduction}

\IEEEPARstart{C}{ybersecurity} is a high and ever-growing priority for many organizations and country governments around the world. Across multiple industries, companies have been moving increasingly towards digitization of records and data, as well as increased the number of digital suppliers they rely on and the overall complexity of their digital supply chain. These trends have coincided with a rapid growth in cybersecurity incidents. Many of these incidents result in data breaches, which correspond to any form of unauthorized access to privately held data with potential compromise of its confidentiality \cite{Sen2015}. The Global Risks Perception Survey conducted by the World Economic Forum in 2019 \cite{WorldEconomicForum2019} ranks “massive data fraud and theft” as the fourth important global risk by likelihood over a 10-year horizon. The IBM and Ponemon 2019 report \cite{IBM2019} estimates  that the average total economic cost of a single data breach incident for an organization is \$3.92M worldwide and \$8.19M in the US, with long-term impacts on companies’ reputation, products and clients’ trust. Cyberattacks on government organizations can be catastrophic, allowing access to confidential information and potentially affecting important government functions.

Various recent federal and industry initiatives and directives in the US aim at improving cybersecurity as a risk management discipline, including the 2021 White House Executive Order \cite{ExecOrder}, as well as the 2022 proposed rules by the US Security and Exchange Commission on Cybersecurity Risk Management, Strategy, Governance, and Incident Disclosure by Public Companies \cite {SEC}. Reliable risk assessment is perhaps one of the most important elements to better manage cybersecurity risks.

The 2020 Data Breach Investigation Report \cite{Nathan2020} indicates that an enterprise’s cybersecurity is commonly assessed primarily if not entirely based on its internal resilience with respect to multiple attack vectors and modalities, such as phishing or spam, and sources of attack, such as an Information Technology (IT) vulnerability or misconfiguration. Some newer approaches to assess cybersecurity risk levels rely on more quantitative methods, such as leveraging externally observable cybersecurity performance data to create cybersecurity risk ratings (referred to hereafter as \textit{cybersecurity ratings}). These ratings capture a range of internal parameters of the enterprise and its IT system (e.g. software versions, open ports, etc.) that can be observed by outside parties. 

However, there is an increasing number of cyberattack incidents, including multiple high profile ones, where perpetrators gained access to an organizations' assets by exploiting weaknesses in their supply chain partners and the overall complexity of the digital supply chain, including recent attacks involving Solarwinds \cite{SolarWindsNPR} \cite{SolarWindsTech}, Kaseya \cite{Kaseya}, Okta \cite{Okta}, Log4J \cite{Log4JGuide} \cite{Log4JTech}, Entrust \cite{Entrust}, Github \cite{GitHub}, Onus \cite{Onus}, NHS \cite{NHS}, Facebook \cite{Facebook}, and Capital One \cite{CapitalOneCNN} among many others.

As already mentioned, current risk assessment methodologies primarily focus on internal infrastructure and processes, such as IT systems and security procedures and protocols \cite{Sen2015}. For example, the Digital Forensic Readiness model \cite{Elyas2015} focuses mainly on the robustness of internal security protocols and IT and control infrastructure like Medical Record Systems \cite{Chernyshev2019}. Other models, such as the one described in \cite{McLeod2018} attempt to expand this focus to consider factors related to internal business processes, such as ones ensuring compliance with data security practices (e.g., the Health Insurance Portability and Accountability Act in the healthcare sector), or to other internal organizational factors, such as the number of employees and facilities. Further evidence-based analysis has been conducted on how following best practices, as outlined by the NIST cybersecurity framework, can correlate to avoidance of data breach incidents \cite{boyson2022}. However, as illustrated by the incident examples discussed above, relying merely on internal features is not likely to be sufficient to holistically capture the respective complex cyberattack risk drivers. 

Data driven risk analysis models and approaches have been developed in the context of physical supply chains. Baryannis et al. and Yang et al. provide overviews of the application of artificial intelligence and machine learning, respectively, in recent supply chain risk  management research and propose future directions \cite{baryannis2019} \cite{yang2023}. Simchi-Levi et al. introduce a data-driven modelling approach to evaluate the impact of disruptions on a global automotive supply chain \cite{simchilevi2015}. The work in this paper expands upon these prior studies by being the first to use large-scale data-driven modeling and analysis of the digital supply chain attributes to assess and identify disruptive cyber event risk drivers.

This study leverages a unique self-constructed comprehensive dataset of more than 30,000 diverse companies, referred to hereafter as \textit{entities}, with their respective organizational characteristics, their cybersecurity ratings, their digital suppliers and their data-breach history. As part of the analysis, the entity's digital supply chain network is mapped, and then network science concepts and machine learning methods are used to develop data-driven descriptive and predictive analyses, supply-chain features, and models to capture hypothesized cyber risk drivers. 

This paper is the first to develop a data-driven framework to rigorously assess risks of cyber data breaches. In particular, the paper highlights the fact that the structure and attributes of the enterprise's digital supply chain play a significant role as predictors and drivers of cyber risk. The analysis results may serve as a foundation for improved methods for assessing cyber risks of organizations as well as for more effective countermeasures for mitigating these cyber risks.

\section{Data}

The dataset used in this paper consists of integrated data from Bitsight \cite{BitSight}, a provider of cybersecurity analytics and cybersecurity ratings based on externally observable performance data, and the Veris Community Database \cite{Veris}. The dataset includes 38345 companies that comprised all enterprises from the Health Care (21093), Oil and Gas (9282), and Retail (7970) sectors within the Bitsight database.

\subsection*{Entity data}

For each entity company, the dataset contains information on baseline features, including employee counts and industry sector, as well as the respective monthly cybersecurity ratings from May 2017 to May 2019 for that entity. These cybersecurity ratings are generated by Bitsight based on publicly observable and measurable information, including features such as Patching Cadence (PC), indicating compliance with system, network and application software updates and patching of security vulnerabilities, Sender Policy Framework (SPF), indicating the existence of an email authentication method specifying mail server authorization, Spam propagation, for unsolicited commercial or bulk emails, and others (see SI Appendix for more details of all the observable features utilized by Bitsight). Each of these ratings has its own underlying calculation methodology and range in numerical value from 300 to 820, with higher rating values meaning to represent better security. 


\subsection*{Supply chain data}

Based on publicly observable and measurable information, the dataset includes a comprehensive description of the digital supply chain relationships for each entity company. Digital supply chain connections were constructed using Domain Name System (DNS) records, website link presence, web records, publicly disclosed information, and additional data analytics. Companies that comprise the digital supply chain of an entity company include hardware and software suppliers that directly provide products or service to the entity company. These digital suppliers are called hereafter \textit{third-party suppliers (or third parties)}. Similarly, suppliers of third-party suppliers are called hereafter \textit{fourth party suppliers (or fourth parties)}.

Considering the \textit{global supply chain graph} induced by the dataset, there are overall 38345 entities, 4875 third parties and 5365 fourth parties. Each distinct company corresponds to a \textit{node} in the global supply chain graph. The dataset also includes information about the service or product provided through each connection between an entity and its third parties, as well as between a third-party supplier and the respective fourth parties. These connections are captured as product edges on the induced global supply chain graph.

There is a total of over 600,000 distinct product edges between the entities in the dataset and their third parties and over 8,000,000 product edges between third parties and respective fourth parties. The product edges are further classified into \textit{product types}. There are over 70 different product types and they include, for example, server technologies, DNS services, and payment processing. Finally, for each third-party and fourth-party supplier, the dataset includes information similar to the entity enterprises, including its own respective outside-in cybersecurity ratings, employee count and industry sector (see SI Appendix for more details). 

Most importantly, for each entity, the dataset captures its \textit{local supply chain} that is comprised of the entity itself, its third parties, and its fourth parties. Notably, there are local supply chains in which a supplier could provide service directly to the entity and also to one of the entity's suppliers. In such cases, this supplier would be considered as only a \textit{third-party} supplier within that particular local supply chain. These suppliers may be considered to be fourth parties in other local supply chains, where they merely act as a supplier to a supplier of the respective entity with no direct supplier relationship with the entity. This is motivated by the fact that from a "cyber attacker perspective", direct connections are generally a preferable vector. This assumption implies that the respective local supply chain network graphs do not include cycles.

\subsection*{Breach incident data}

The dataset also integrates extensive data on documented breaches from May 2017 through May 2020 from data obtained from Bitsight \cite{BitSight} and the Veris Community Database \cite{Veris}. The dataset compiles publicly available information on companies that were breached. Overall, the dataset includes 8,344 unique breach incidents. Out of all the entities in the dataset, 3.29\% had experienced a breach incident during the three years of the study period. Specifically, 1.00\% had a breach incident during May 2017 to May 2018, 1.37\% during May 2019 to May 2020, and 1.55\% during May 2019 to May 2020. The increasing proportion over each year period can be explained by at least two underlying factors. The first factor is that cyber data breach events are becoming more common for enterprises as time passes, either as a result of improved attack strategies or increase in the number of overall attacks. The second factor relates to potential gradual improvements in the reporting consistency and documentation of data breach incidents.


\section{Analysis}

The paper first provides a descriptive analysis of the global supply chain network structure described above. This motivates the introduction of the notion of the entity's \textit{local} supply chain network, consisting of the entity's third-party suppliers and their respective fourth-parties. The local supply chain network is then further analyzed to develop original features based on the local supply chain structure and attributes, hypothesized to be associated with cyber data breach risk. These features are used to develop and evaluate a machine-learning predictive risk model to assess the role of the local supply chain features in predicting risk of future cyber data breaches.

\begin{figure}[!t]
    \centering
    \includegraphics[width = 0.5\textwidth]{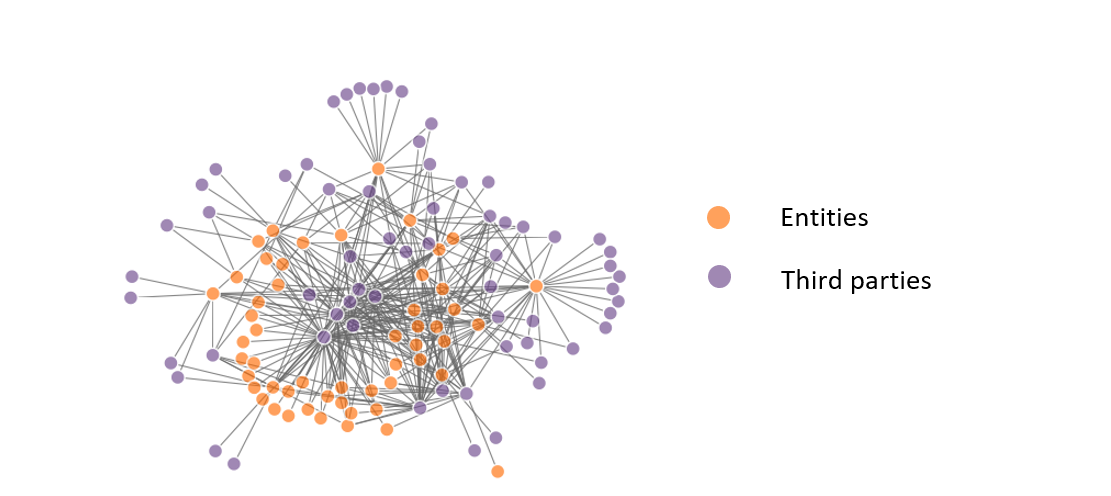}
    \caption{Network Visualization of a subset of retail entities and third parties }
    \label{fig:network}
\end{figure}

\subsection*{Global supply chain network visualization and description}

To gain insights with respect to how supply chain structure and attributes can play a role in driving cyber data breach risk, the sector-specific (e.g., Retail or Oil and Gas) global supply chain networks are created and visualized. For ease of interpretation, only entities and their respective third-party suppliers are included. The analysis is focused on the distribution of the graph degree of the third-party suppliers, which corresponds to the number of customers they serve.

In order to understand how supply chain network structure of the entities and their digital suppliers can be a source of cyberattack risk, the sector-specific (e.g. Retails or Oil and Gas) supply chain networks are created and visualized, consisting of the respective entities and their third-party suppliers. Fourth parties are excluded for ease of interpretation. Additionally, it defines each supplier's degree as the number of customers they serve, and compares the customers served by large and small degree suppliers. This first step aims to identify structural patterns in the supply chain network and potential sources of differentiated cyber risk.

The global supply chain network of the entities in the retail sector includes 7970 entities and 3156 third-party suppliers. Figure \ref{fig:network} provides a network visualization using a randomly sampled subset of the retail entities within the dataset as well as their most important third-party suppliers (see Appendix for supplier importance details). Specifically, Figure \ref{fig:network} captures a network with a total of 117 nodes including 53 retail entity companies, 64 third parties and 370 edges. it can be observed from Figure \ref{fig:network} that some of the third parties have near universal connectivity to almost all of the entities within the supply chain network. These are typically large service providers with huge customer bases, and they are a source of common risk to many entities. That is, compromising these suppliers could facilitate a cyberattack and potential data breach of many entities simultaneously.


Additionally, Figure \ref{fig:network} also includes third-party suppliers with low-connectivity. These tend to supply mostly highly connected entities (i.e., ones who receive service from relatively many suppliers). Furthermore, these are a source of residual (or differentiated) risk for the specific entities that they serve. That is, compromising these suppliers will only allow potential access to the small number of entities that they serve. The more suppliers with low connectivity an entity has, the higher its individual residual risk might be (see Appendix for additional visualizations). 

\begin{figure}[!t]
    \centering
    \includegraphics[width  = 0.5\textwidth]{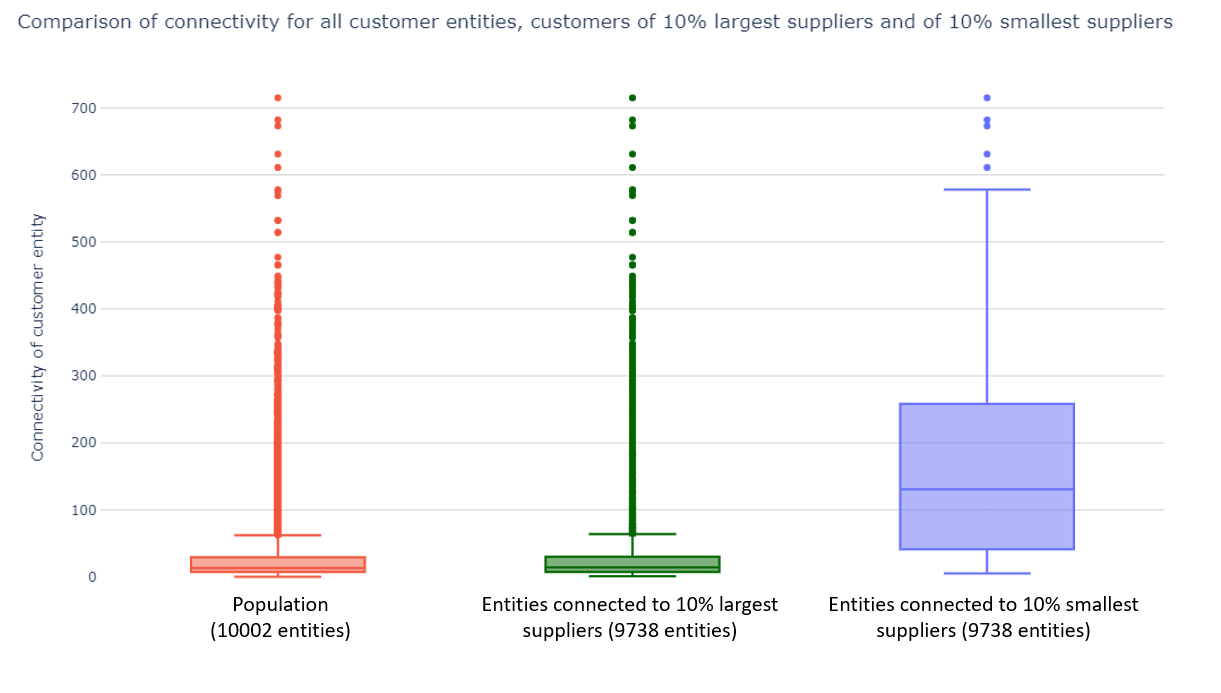}
    \caption{Comparison of supplier connectivity for all entities and those served by large and small suppliers. Large suppliers are those with the top 10\% highest supplier degree while small suppliers are those with the bottom 10\% supplier degree. }
    \label{fig:connectivity_distributions}
\end{figure}

These insights are further confirmed by degree distribution analysis on the entire retail sector supply chain graph that is presented in Figure \ref{fig:connectivity_distributions}. Specifically, considering the third-party suppliers with highest and lowest connectivity (top 10\% and lowest 10\% graph degree among third parties, respectively), the figure contains three plots of the degree distribution (i.e., the number of connections an entity has), including all the entities, the entities served by suppliers with highest connectivity and entities served by suppliers with lowest connectivity. The plots include 10002, 9737, and 310 entities respectively. The degree distributions among all entities and entities served by third-party suppliers with the highest connectivity are nearly identical, with respective median degrees of 13 and 14. In contrast, the degree distribution of entities served by third-party suppliers with the lowest connectivity is significantly higher, having a median of 131. This difference is statistically significant on \textit{t}-tests, having p-values of essentially 0.


\subsection*{Local supply chain network features}

To assess the individual risk level of a specific entity, the analysis attempts to characterize the attributes of the entity's local supply chain network defined above using original supply chain features. This is motivated and informed by the increased prevalence of supply-chain related data breach incidents and also by network risk assessment methodologies \cite{Lewis}. The supply chain features consist of connectivity, product/service type, historical cyberattack, and network-wide outside-in cybersecurity rating features.
    
\begin{figure}[!t]
    \centering
    \includegraphics[width = 0.5\textwidth]{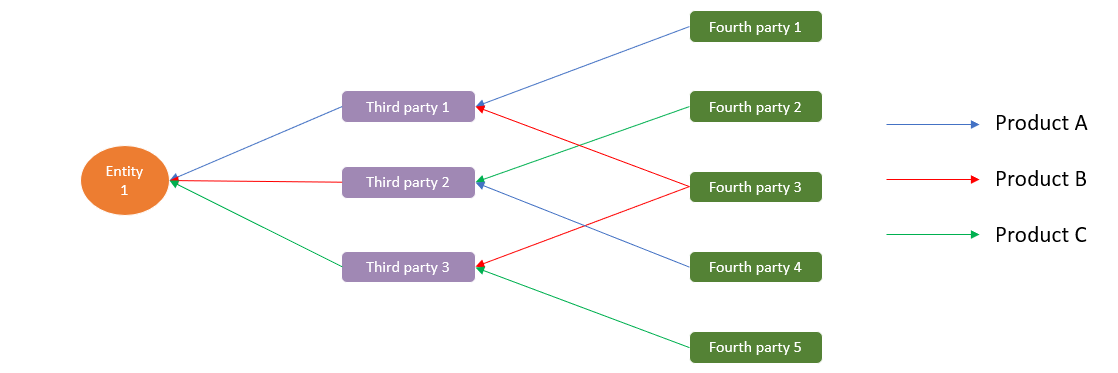}
    \caption{Third and fourth party connectivity. This figure illustrates connectivity features design. Only one sector of third parties and one sector of fourth parties are represented. Here, third-party connectivity for this sector is equal to 3 while fourth party connectivity, which is the median of third-party connections to fourth parties, is equal to 2. }
    \label{fig:connectivity_design}
\end{figure}
    
Connectivity features describe the number of suppliers within the local supply chain network belonging to each sector. As illustrated in Figure \ref{fig:connectivity_design}, sectorial third-party connectivity indicates the number of third parties for each sector. Similarly, sectorial fourth party connectivity defines the number of fourth parties of each sector in the local supply chain. The study then defines a supply chain network exposure feature that corresponds to the total number of distinct third and fourth parties of an entity. This feature enumerates the number of potential sources of supply chain breach propagation, either from third or fourth parties, that can occur within the local supply chain network.

Network-wide summary features are developed using the mean outside-in cybersecurity rating values of third and fourth parties in the network to approximate the average internal cybersecurity level of the local network. Additional summary features utilizing third and fourth party employee counts are also developed to capture the average size of suppliers within the local network. These features aim to provide 

Additionally, an aggregated edge-based supply chain network feature is designed utilizing the various types of supplied products being supplied to the entity. Figure \ref{fig:connectivity_design} demonstrates the product-edge relations that are present between suppliers and customers in the local supply chain. To construct the aggregate feature, each product is ordered according to its relative risk level (see Appendix for risk level computation details). Based on this ordering, the products are sorted into 4 quantiles corresponding then to 4 risk groups. For each entity, the number of edges belonging to each of the 4 groups are counted and then aggregated via a weighted sum to create the product-edge aggregate feature, where a weight of 4 is given to the highest risk quantile, 3 to the second highest, 2 to the third highest, and 1 to the lowest risk quantile.


Finally, for each entity, the number of breaches that had been reported 2 years prior to May 2019 by the third parties were counted to describe the historical vulnerability of the company's direct suppliers. A similar feature was constructed using breaches reported by the fourth parties for indirect suppliers. This feature is motivated by the fact that supply chain related cyberattacks typically start with the attacker compromising a supply chain partner of the attack's target (see Appendix for a table of all of these network features and their descriptions).

All of these features of interest can potentially provide insights about the overall attacker access and security of the supply chain network. Additionally, these supply chain network features may also capture the inherent complexity of the entity company's digital infrastructure. Companies that need to manage highly complex digital infrastructure might be more vulnerable and at a higher risk. All of these hypotheses motivate the development and inclusion of local supply chain network features in the predictive models introduced in the subsequent section.

\subsection*{Machine learning framework} 

To assess the predictive value of the supply chain network features, three machine learning models are trained on the data and evaluated out of sample. Model 1 (baseline) includes only basic features on each entity, specifically, sector and the number of employees. Model 2 adds to that the entity's outside-in cybersecurity rating features. Finally, Model 3 includes in addition the local supply chain network features.


All the models are trained using the binary classification gradient boosting algorithm \cite{Agarwal1994} which integrates interaction between features and delivers the highest out-of-sample predictive power of the models tested. A stratified split of the dataset into a training (70\%) and test (30\%) set is performed to get a balanced split with respect to sector and data breach status. Then, a 5-fold cross-validation is applied on the training set for hyper parameter tuning to define the best learning rate, number of estimators, maximum depth and L1 regularization parameter \cite{refaeilzadeh2009cross}. This procedure is repeated on 20 random training-test splits to confirm the consistency of the performance assessment. 

The output of each of the models is a risk score for each entity.

\begin{table*}
\centering
\caption{Models out-of-sample performance results}
\begin{tabular}{cc}
 & AUC \\
1. Basic features & 76.5\%  \\
95\% confidence interval & 76.1\%-76.8\%\\
\hline
2. Basic features and entities cyber scores  & 77.8\%  \\
95\% confidence interval & 77.3\%-78.2\%\\
\hline
3. Basic features, entities, suppliers cyber scores and supply chain structure	& 78.8\%  \\
95\% confidence interval & 78.4\%-79.3\% \\
\end{tabular}
\\
Table 1 summarizes the out-of-sample AUC of all three models with their 95\% confidence intervals
\label{tab:model_performance}
\end{table*}


\section{Machine learning cyber risk assessment}

Table \ref{tab:model_performance} summarizes the respective out-of-sample performance of Models 1, 2, and 3 using the area under the curve (AUC) metric to evaluate model performance with a 95\% confidence interval over 20 trials. In these results, the basic model with employee count and sector achieves an AUC of 76.5\%. When the Bitsight score features are included in Model 2, the AUC increases to 77.8\%. The AUC of Model 3 further increases to 78.8\% upon adding the local supply chain features. Both of these AUC increases are significant at the 95\% confidence level.


The performance of Model 1 implies that organizational features are predictors of cyber risk, as the model already outperforms a random guess AUC of 50\%. This supports the idea that cyberattacks are not uniform among all companies, but instead are more likely to target large organizations and certain sectors, like healthcare, that have more attractive data and could be more vulnerable. The improvement in AUC from Model 1 to Model 2 provides evidence that outside-in cybersecurity ratings are effective in providing additional useful information about the internal cyber risk posture for individual entities that cannot be derived from the baseline features.

Model 3, which includes the local supply chain features underscores the added value of considering an entity’s digital supply chain cyber history and structure to assess the risk of future breaches. The improved performance supports the hypothesis that supply chain attributes provide an important signal regarding the cyber data breach risk level of an entity beyond what internal features are able to capture. The predictive power of digital supply chain features could be related to the observed increasing number of cyberattacks that are conducted through third-party suppliers. More generally, considering supply chain features can provide additional, and potentially more accurate, assessments of the overall enterprise IT processes and system complexity that in turn could correlate with the vulnerability of the enterprise to cyber data breaches.

\begin{figure}[!b]
    \centering
    \includegraphics[width = 0.5\textwidth]{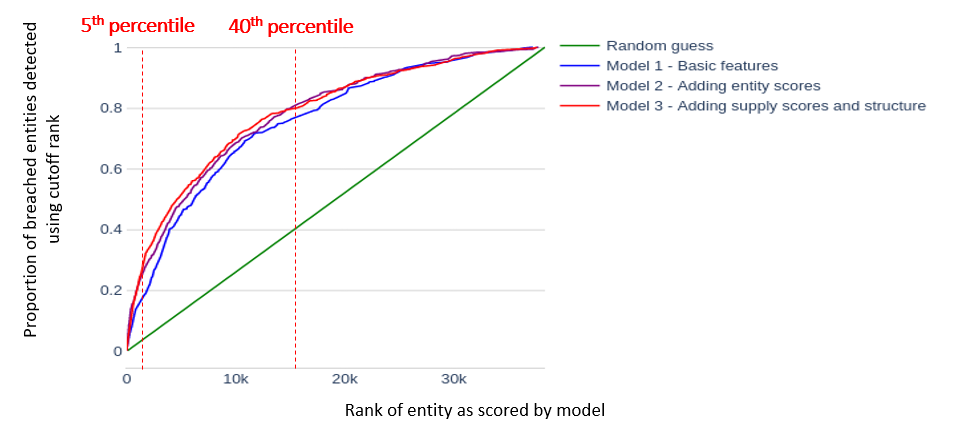}
    \caption{Comparison of the detection score, and their rank from their out-of-sample cybersecurity score based on a random guess, Model 1, and Model 2.}
    \label{fig:rank_comparison}
\end{figure}

\section{Detection rate comparison}

Another metric to further compare the relative predictive power of the three models is the \textit{detection rate}, which is calculated as follows. For each model, the entities in the test data are ranked from highest to lowest model score. The detection rate is calculated based on the proportion of breached entities identified  for any specified number of top ranked entities.

Figure \ref{fig:rank_comparison} plots the out-of-sample detection rate of each of the models. A random guess performance is also included to serve as a benchmark comparison. A perfect predictive model would rank all the entities that were breached at the top of the list. The figure reveals that all three models have a significantly better detection rate than a random guess. Both Model 2 and Model 3 perform noticeably better than Model 1 except for the detection rate among the highest ranked entities. Additionally, Model 3, which uses both supplier cybersecurity ratings and supply chain features, has noticeably better detection rate than Model 2 between the $5^{th}$ and $40^{th}$ percentiles of the ranked companies.

\section{Feature assessment for identifying cyber risk drivers}

\begin{figure}[!b]
    \centering
    \includegraphics[width = 0.5\textwidth]{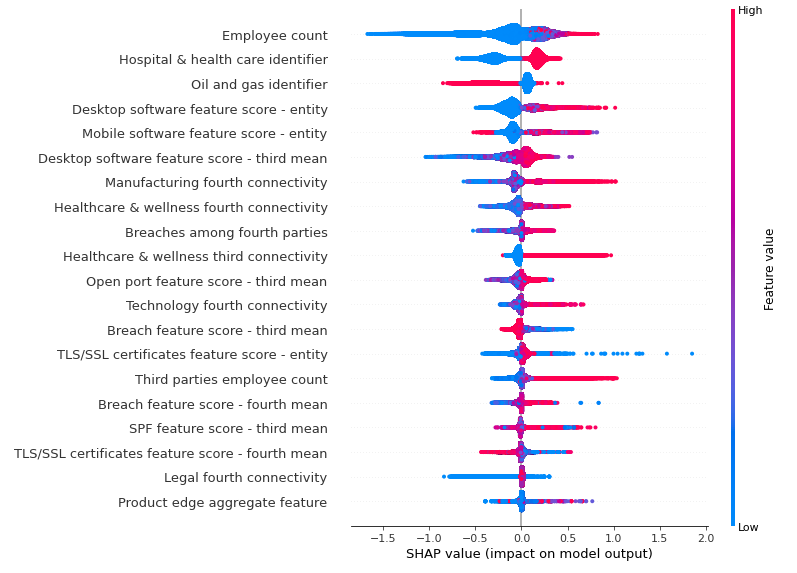}
    \caption{Features importance and impact. Shapley impact is computed through the python SHAP library based on \cite{Lundberg2019}. "Feature value” corresponds to its numerical relative value. }
    \label{fig:feature_impact_interaction}
\end{figure}

To identify potential cyber risk drivers for breached entities, the most important features utilized by Model 3 are examined. Figure \ref{fig:feature_impact_interaction} (A) illustrates the Shapley impact of the top 20 features with highest importance values in Model 3 \cite{Lundberg2017a, Lundberg2019}. 

The most important organizational risk drivers are the \textit{Hospital and healthcare identifier} and employee count, for which the larger the values are in red, the riskier a company is considered. The \textit{Oil and gas identifier} on the other hand shows a negative correlation with risk in the model. 

The following local supply chain features appear to be predictive of increased cybersecurity risk. At the third-party level, there are the Desktop software, Open port, Breach, and SPF third-party summary features. These features represent the mean of the third parties' outside-in cybersecurity rating values for each respective rating. Additionally, there is the third party employee count feature. This feature is computed using the average employee count of third parties connected to the entity. Connectivity to Healthcare and wellness third parties is also among the top 20 risk drivers. This feature counts the number of direct Healthcare and wellness suppliers an entity has. Finally, the product edge aggregate feature is also one of the top 20 risk drivers. This feature is computed as the weighted sum of the number of products from each risk group that the entity received. (See SI appendix for a full description of all features presented in Figure \ref{fig:feature_impact_interaction})

At the fourth party level, there are the Manufacturing, Healthcare and wellness, Technology, and Legal fourth party connectivity features. These features count the number of fourth parties connected to the entity from each respective sector. Secondly, there is the breaches among fourth parties feature. This feature captures the number of historical breaches that were experienced by fourth parties in the local network. Finally, there are the Breach and TLS/SSL certificates fourth-party summary features. These represent the mean of the fourth parties' outside-in cybersecurity rating values for each respective rating.

The result regarding the healthcare feature is in line with the current literature, as it points out the healthcare sector as one of the \textit{most susceptible to publicly disclosed data breaches} \cite{Abouelmehdi2017b}. Reasons for this could be the value of sensitive patients’ information, the nature of the industry that is heavily data driven and progressively digitizing \cite{Abouelmehdi2017b}, and the reporting requirement of these companies as a result of HIPAA security regulations.  Additionally, the importance of employee count could be explained by increased individual vulnerabilities and risky behaviors from a cyber perspective that result in social engineering incidents, employee data theft and data leaks as explained in \cite{Hadlington2017}. 

The model also reveals system vulnerabilities at different levels of the supply chain that could impact an entity's cybersecurity. Connectivity to fourth parties of various sectors may correlate to entity cyber risk because they introduce potential differentiated risk avenues to access the entity. It is possible that more fourth parties increase the number of vulnerable entry points to the local supply chain which an attacker can leverage in order to then penetrate the entity itself. The importance of the number of breaches among fourth parties further supports the idea that having fourth parties that are vulnerable to data breach incidents can correlate to increased entity cyber risk.

The third-party cybersecurity rating measures serve as proxies for third-party supplier security maturity and may capture additional risk information regarding secure connections between entities and their suppliers. In other words, these ratings could represent information not only about whether an attacker has greater ease in accessing the third parties, but also if it is easy to gain access to their customers' systems. Connectivity to large third parties may be correlated to data breach risk because these third parties are attractive targets for cyber attackers, which could affect customers downstream. Healthcare and wellness third-party connectivity features are critical not only as an indicator of health care parent entities but also as a proxy for the number of their third-party suppliers.

Finally, the high importance of the product aggregate feature indicates that companies with a greater number of products and services received are at greater risk of being breached, potentially because a high number of products received could imply that an attacker has more opportunities to identify supplier-to-entity connections that will allow an attack on the entity's IT system.

It is also important to note that these local supply chain features could be further capturing information regarding an entity's overall complexity. As such, they could provide significant signals about an entity's vulnerability to non-third-party related breach incidents as well.

\begin{figure}[!b]
    \centering
    \includegraphics[width = 0.5\textwidth]{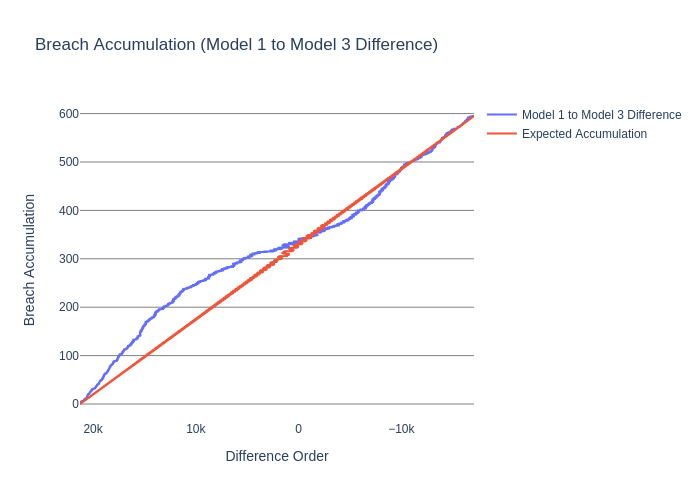}
    \caption{Comparison between observed accumulated breached entities when ordering by rank difference between Model 3 to Model 1 and expected accumulated breached entities using a random ordering. X-axis adjusted to show where difference is positive or negative.}
    \label{fig:breach_1_3}
\end{figure}

\section{Analysis of entities with high-inter-model risk ranking differences}

\begin{figure}[!t]
    \centering
    \includegraphics[width = 0.5\textwidth]{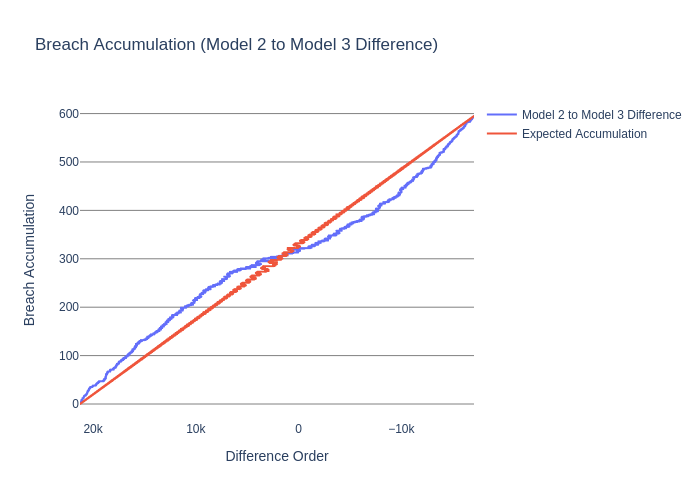}
    \caption{Comparison between observed accumulated breached entities when ordering by rank difference between Model 3 to Model 2 and expected accumulated breached entities using a random ordering. X-axis adjusted to show where difference is positive or negative.}
    \label{fig:breach_2_3}
\end{figure}

To further understand the performance differences between Model 3 and Models 1 and 2, respectively, for each pair of models (3-1 and 3-2), the entities are ordered according to the rank difference between the two respective models from highest to lowest. Then the cumulative number of companies that were breached as one moves down the ordered list of entities is counted. This is then compared to the expected number of accumulated breached entities if using a random ordering instead, which is represented as a straight line in Figures \ref{fig:breach_1_3} and \ref{fig:breach_2_3}.



In Figure \ref{fig:breach_1_3}, the breach accumulation for entities with a positive difference is much higher than the expected breach accumulation. In fact, there is a visible separation when considering entities that have a positive difference in rank. An additional paired \textit{t}-test between the accumulated breaches for entities with a positive difference and the expected accumulated breaches confirms that this difference is significant with a p-value of essentially 0. Figure \ref{fig:breach_2_3} demonstrates a similar pattern. 

The patterns displayed within these two graphs reveal that the entities that the supply chain model rates as riskier compared to the other two models appear to be significantly more likely to be breached. Thus, this serves as additional evidence that the inclusion of supply chain features allows the model to more effectively identify vulnerable companies.




\begin{table*}[h!]

\caption{Top 10 Positive Difference Entities from Model 3 to Model 1 with Percentiles (Most Risky)}

\begin{tabular}{p{2cm}ccccccc}
Entity rank & Employee count & Fourth party breaches & Product Edges & Healthcare third connectivity & Manufacturing fourth connectivity\\

Mean & 4143 & 44.79 & 461 & 0.12 & 14.22  \\
\hline
1 & 100 (34.18\%) & 100 (99.84\%) & 5900 (98.82\%)& 2 (98.27\%) & 34 (99.91\%)\\
\hline
2 & 32 (0.10\%) & 90 (99.81\%) & 27635 (99.77\%)& 1 (96.09\%) & 31 (99.36\%)\\
\hline
3 & 100 (34.18\%) & 71 (98.54\%) & 8243 (99.15\%) & 1 (96.09\%) & 28 (97.61\%)\\
\hline
4 & 100 (34.18\%) & 69 (98.36\%) & 7947 (99.11) & 1 (96.09\%) & 26 (95.54\%)\\
\hline
5 & 100 (34.18\%) & 76 (97.18\%) & 1040 (94.27)& 2 (98.27\%) & 31 (99.36\%)\\
\hline
6 & 100 (34.18\%) & 96 (99.74\%) & 19910 (99.65\%) & 14 (99.97\%) & 30 (98.94\%)\\
\hline
7 & 100 (34.18\%) & 75 (96.80\%) & 1625 (96.07\%) & 2 (98.27\%) & 29 (98.35\%)\\
\hline
8 & 100 (34.18\%) & 75 (94.22\%) & 1625 (96.07\%)& 2 (98.27\%) & 29 (98.35\%)\\
\hline
9 & 100 (34.18\%) & 76 (99.42\%) & 220 (81.83\%)& 1 (96.09\%) & 22 (88.79\%)\\
\hline
10 & 10 (4.88\%) & 85 (99.93\%) & 6016 (98.85\%)& 0 (47.21\%) & 31 (99.36\%)\\

\label{tab:most_risky}
\end{tabular}
\\
Table 2 records several feature values for the 10 entities that had the most positive rank difference from Model 3 to Model 1.
\\[0.25in]


\caption{Top 10 Negative Difference Entities from Model 3 to Model 1 with Percentiles (Less Risky)}
\begin{tabular}{p{2cm}ccccccc}
Entity rank & Employee count &  Fourth party breaches & Product Edges & Healthcare third connectivity  & Manufacturing fourth connectivity  \\
Mean & 4143 & 44.79 & 461 & 0.12 & 14.22\\
\hline
1 & 30000 (96.88\%) & 42 (38.25\%) & 40 (40.86\%)& 0 (47.21\%) & 13 (36.35\%)\\
\hline
2 & 209746 (99.85\%) & 14 (4.92\%) & 3 (2.22\%)& 0 (47.21\%) & 0 (1.13\%)\\
\hline
3 & 30000 (96.88\%) & 46 (51.96\%) & 75 (61.48\%) & 0 (47.21\%) & 14 (43.33\%)\\
\hline
4 & 30000 (96.88\%) & 53 (71.62\%) & 657 (91.86\%) & 0 (47.21\%) & 19 (78.00\%)\\
\hline
5 & 32000 (98.24\%) & 15 (5.76\%) & 20 (20.96\%)75  & 0 (47.21\%) & 4 (11.05\%))\\
\hline
6 & 345000 (99.94\%) & 42 (38.25\%) & 28 (29.04\%)& 0 (47.21\%) & 12 (29.76\%)\\
\hline
7 & 239754 (99.87\%) & 41 (34.78\%) & 14 (14.90\%)& 0 (47.21\%) & 13 (36.35\%)\\
\hline
8 & 50005 (98.85\%) & 39 (28.71\%) & 47 (46.67\%)& 0 (47.21\%) & 11 (25.86\%)\\
\hline
9 & 500000 (99.97\%) & 32 (19.16\%) & 30 (31.06\%)& 0 (47.21\%) & 9 (22.28\%)\\
\hline
10 & 70000 (99.20\%) & 30 (17.03\%) & 93 (67.00\%)& 0 (47.21\%) & 10 (24.22\%)\\
\label{tab:least_risky}
\end{tabular}
\\
Table 3 records several feature values for the 10 entities that had the most negative rank difference from Model 3 to Model 1.
\end{table*}

\section{Analysis of distinguishing features that explain inter-model rank differences}

To analyze the features that explain the rank differences between Model 3 and Model 1, the 10 entities with the most positive and negative rank differences are analyzed. Specifically, the values of the baseline features as well as important supply chain features (according to the Shapley impact value) are then analyzed. Tables \ref{tab:most_risky} and \ref{tab:least_risky} contain the employee count, second degree breach, product edge, healthcare third party connectivity, and manufacturing fourth party connectivity features for each of these entities, as well as the percentile for these features relative to the rest of the population. Additionally, the mean value across the population for each of these features is included.

Table \ref{tab:most_risky} presents the entities with the largest rank difference from Model 3 to Model 1. Notably, all the entities in the table have an employee count of 100 or less, which is far below the mean. In comparison, nearly all of the supply chain related features for each of these entities are within the top 95\textsuperscript{th} percentile of the population. 

Table \ref{tab:most_risky} highlights the fact that Model 1 relies heavily on employee count information to produce its predictions and thus rates most companies with low employee counts, such as the ones included in Table \ref{tab:most_risky}, as safe. Model 3 is instead able to differentiate further utilizing the local supply chain network features. As such, entities that are small according to employee count but have large local supply chain network features such as the ones included in the table are rated as far more relatively risky by Model 3 compared to Model 1.

Conversely, the entities in Table \ref{tab:least_risky}, which have the most negative rank difference, have larger employee counts, with all of them having an employee count greater than the mean and all but one being in the 95\textsuperscript{th} percentile. On the other hand, the local supply chain network feature values for these entities are almost all below the overall population mean. Although these large entities are being rated as risky by Model 1 because of their employee count, Model 3 differentiates them from other large entities by using their smaller supply chain network features to rate them as far less relatively risky (See SI appendix for a similar analysis conducted between Models 3 and 2).

While not exhaustive, this analysis provides evidence for the important role that local supply chain network features play in how Model 3 is differentiating its risk assessment from the other two models.

\section{Discussion and Conclusion}



This paper establishes the significance of data-driven analysis for deriving important and actionable new cyber risk insights. It demonstrates that a company’s supply chain structure and attributes provide an important signal in assessing its cyber risk of future data breaches. Indeed, it provides evidence that relying only on indicators related to a company’s internal cybersecurity management may not capture important risk predictors. As the predictive modelling approach and additional models' ranking analyses have demonstrated, utilizing the digital supply chain information can significantly improve the ability to more successfully assess the cybersecurity risk of a company. The assessment of the features' relative importance further highlights the relevance of the local supply chain features in assessing and predicting risk, particularly third-party connectivity, fourth-party connectivity, and the overall security of the suppliers within a company's direct or indirect network.

The findings in this paper provide important insights for practice as well as for future research of cybersecurity risk assessment and management. 

First, identifying important predictors of cyber risk related to the enterprise and its supply chain can significantly improve the ability to reliably and proactively assess and manage risk, and it informs what risk indicators should be monitored. The paper's analyses suggest that while it is important to have comprehensive enterprise cybersecurity internal policy controls and procedures, an effective supply chain cybersecurity risk assessment and mitigation system could have significant additional benefits. In particular, current cybersecurity risk assessment methodologies need to be enhanced to include supply chain network-related indicators to more comprehensively characterize and manage related risks. This will consequently drive more effective cybersecurity decisions and countermeasures.

Secondly, the paper also stimulates and highlights important further research directions. The local supply chain network construction and feature design developed in this study can inform future work for assessing cyber risk within supply chains and large network structures in general. The importance of local network connectivity features and the product edge feature as risk drivers motivate potential research to explore their role in the mechanisms of cyberattack propagation within supply chain networks. Moreover, increasingly comprehensive and complex features could be built from the overall network structure, which may prove to be even more successful in predicting data breach and cyberattack occurrences. This paper focuses on the digital supply chains of the respective entities, but more research is needed to understand to what extent the structure and dynamics of physical supply chains are predictive of cyberattack risk. Additionally, future work can expand upon the data sources used and design complementary supply chain network features. Finally, the results in the paper motivate the need to standardize how data breach and cyberattack incidents are documented and categorized, especially when it comes to differentiating attacks that are related to the supply chain. More generally, it would be important to develop a comprehensive data breach and cyberattack taxonomy that properly distinguishes cyber risks originating from a company's internal system versus its external supply chain. 

Ultimately, such data-driven approaches will contribute to the important challenge of improving the management of current and future cybersecurity risks with more effective visibility, assessments, insights, and countermeasures.

This study also has several limitations that should be noted. First, it should be noted that the data on cyberattacks and data breaches is censored by inconsistent reporting, and the level of reporting could have changed over time. This paper utilizes both private and publicly available data, but future efforts to share data across industry stakeholders could improve the quality and consistency of the respective data breach incidents data, and would allow the development of improved models. Second, while outside-in cybersecurity ratings provide an approximate signal regarding a company's internal cybersecurity resilience, these methods are not perfect, and they are limited to publicly observable data that do not capture many aspects related to the internal cybersecurity posture of a company. Additionally, these ratings are proprietary and do not necessarily amount to a unified risk assessment framework that can be shared across an industry. Again, future effort to expand and standardize across industry stakeholders would be important. Finally, the paper suggests hypothesized risk drivers that seem associated with data breach incidents, but more work is needed to better understand the causal mechanisms and the specific vectors of attack, and what risk indicators best capture them.

\section{Acknowledgements}
We would like to thank Tom Montroy and BitSight staff for providing data and feedback to support this project. The work of the second author was partially funded by CSAIL at the Massachusetts Institute of Technology, and we are grateful for their financial support.

\ifCLASSOPTIONcaptionsoff
  \newpage
\fi




\bibliographystyle{plain}
\bibliography{Cybersecurity_supply_chain}

\vfill


\end{document}